\documentclass[useAMS]{mn2e}
\def\Msun{M$_{\odot}$ }

\input epsf
\title[M82-F: A Doomed Super Star Cluster?]{M82-F: A Doomed Super Star 
Cluster?\thanks{Based in part on 
observations with the NASA/ESA {\it Hubble
Space Telescope}, obtained at the Space Telescope Science Institute,
which is operated by the Association of Universities for Research in
Astronomy (AURA), Inc., under NASA contract NAS 5-26555.  Also based
on observations obtained with the William Herschel and the WIYN
telescopes.}}

\author[L. J. Smith and J. S. Gallagher] {Linda J. Smith$^1$
and John S. Gallagher, III$^2$\\ $^1$Department of Physics and
Astronomy, University College London, Gower Street, London, WC1E 6BT\\
$^2$Department of Astronomy, University of Wisconsin-Madison, 5534
Sterling, 475 North Charter St., Madison WI 53706, USA} 
\date{Accepted 2001 April 23.}

\pagerange{\pageref{firstpage}--\pageref{lastpage}}
\pubyear{2001}

\begin{document}

\maketitle

\label{firstpage}

\begin{abstract}
We present high dispersion echelle spectroscopy of the very luminous,
young super star cluster (SSC) `F' in M82, obtained with the 4.2-m
William Herschel Telescope (WHT), for the purpose of deriving its
dynamical mass and assessing whether it will survive to become an old
globular cluster.  In addition to the stellar lines, the spectrum
contains complex Na\,{\sc I} absorption and broad emission lines from the
ionized gas.  We measure a stellar velocity dispersion of
$13.4\pm0.7$~km\,s$^{-1}$, a projected half-light radius of
$2.8\pm0.3$~pc from archival {\it HST/WFPC2\/} images, and derive a
dynamical mass of $1.2\pm0.1 \times 10^6$~M$_\odot$, demonstrating
that M82-F is a very massive, compact cluster.  We determine that the
current luminosity-to-mass ratio $(L_V/M)_\odot$ for M82-F is
$45\pm13$.  Comparison with spectral synthesis models shows that
$(L_V/M)_\odot$ is a factor of $\sim 5$ higher than that predicted
for a standard Kroupa (2001) initial mass function (IMF) at the
well-determined age for M82-F of $60\pm20$~Myr.  This high value of
$(L_V/M)_\odot$ indicates a deficit of low mass stars in M82-F; the 
current mass function (MF)
evidently is `top-heavy'. We find that a lower mass cutoff of
2--3\,M$_\odot$ is required to match the observations for a MF
with a slope $\alpha=2.3.$ Since the cluster apparently lacks
long-lived low mass stars, it will not become an old globular cluster,
but probably will dissolve at an age of $\leq$2~Gyr.  We also derive
up-dated luminosity-to-mass ratios for the younger SSCs NGC 1569A and
NGC 1705-1. For the first object, the observations are consistent with
a slightly steeper MF ($\alpha=2.5$) whereas for NGC 1705-1, the
observed ratio requires the MF to be truncated near 2~M$_\odot$ for a
slope of $\alpha=2.3$. We discuss the implications of our findings in the
context of large scale IMF variations; with the present data the
top-heavy MF could reflect a local mass segregation effect during the
birth of the cluster.  M82-F likely formed in a dense molecular cloud;
however, its high radial velocity with respect to the centre of M82
($\sim-$175 km~s$^{-1}$) suggests it is on an eccentric orbit and now
far from its birthplace, so the environment of its formation is
unknown.
\end{abstract}
\section{Introduction}\label{intro}
M82 is usually considered the prototype of starburst galaxies since
the luminosity and star formation rate of its central star-forming
region is comparable to the luminosity of a typical giant spiral
galaxy (e.g.  Telesco 1988; Lester et al. 1990). Imaging studies at
optical and infrared wavelengths have shown that young, compact,
luminous star clusters (`super star clusters' or SSCs) are very
numerous in the actively star-forming regions of M82 (O'Connell \&
Mangano 1978; O'Connell et al. 1995; Satyapal et al. 1995, 1997; de
Grijs et al.  2001). M82 provides an excellent and nearby setting in
which to study the properties of young SSCs.  We adopt a distance to
M82 of 3.6~Mpc by assuming that it is at the same distance as M81
(Freedman et al. 1994), consistent with the red giant branch tip
distance to M82 found by Sakai \& Madore (1999).

The presence of luminous, massive SSCs in starburst galaxies leads to
the question: are these really `young globular star
clusters' (e.g. Schweizer \& Seitzer 1993; Whitmore et al. 1993;
O'Connell, Gallagher \& Hunter 1994; Meurer et al. 1995)?  Differences
in cluster luminosity functions between those for young SSCs in
starbursts and those of ancient globular cluster systems may present a
problem.  While globular star cluster luminosity functions peak at $M_V
\approx -7.5$ (corresponding to $M \sim 10^5$\,M$_\odot$; see Harris
1991; Ashman, Conti \& Zepf 1995), the raw luminosity functions of
clusters spawned in starbursts typically resemble the power law
distributions found for open star clusters (e.g. Whitmore et al. 1999;
Zhang \& Fall 1999), although incompleteness may be an issue
(de Grijs, O'Connell \& Gallagher 2001).
A further complication can arise in the evolution
of populations of star clusters within galaxies.  Fritze - v.
Alvensleben (1998, 1999) emphasized the possible importance of
differential survival rates, which favour dense, massive star clusters,
such as compact SSCs, and might transform the power luminosity
functions of star clusters at birth into those of globular clusters 
(see also Chernoff \& Weinberg 1990; Elmegreen \& Efremov 1997;
Gnedin \& Ostriker 1997).
Luminosity function measurements alone for populations of star
clusters do not resolve this issue.

The questions of masses and densities of SSCs are therefore critical;
do SSCs have the mass densities of globular star clusters? This is a
necessary but not a sufficient condition for SSCs to survive for
$\sim10$\,Gyr time scales to become globular clusters. Ho \&
Filippenko (1996a,b) measured stellar velocity dispersions for two
SSCs, NGC 1569A and NGC 1705-1, with the {\sc HIRES} echelle
spectrograph on the Keck I 10-m telescope. Combining these data with
half-light radii in the literature derived from observations with the
{\it Hubble Space Telescope (HST)\/}, they found dynamical masses
consistent with those of globular clusters. Both of these SSCs are
relatively bright ($V=14.8$ and 14.7, respectively), young (ages of $<
20$\,Myr), and in small galaxies with low metallicities.

The recognition that compact SSCs may indeed be proto-globular
clusters opens up the possibility of studying their formation
conditions, evolution and destruction processes.  One issue of
particular interest is whether SSCs have sufficient numbers of low
mass stars to remain as bound systems over long time-scales. The slope
of the stellar initial mass function (IMF)\footnote{There is some
confusion in the literature over the use of the term `IMF' to describe
the observed properties of evolved star clusters. We will use the term
`IMF' for the theoretical initial distribution of masses, and the term
`mass function (MF)' to refer to the present-day mass
distribution of a cluster inferred from observation.} is critical since too
high a fraction of high mass stars will cause the cluster to evaporate
within a few Gyr due to stellar mass-loss (e.g. Chernoff \& Weinberg
1990; Goodwin 1997; Takahashi \& Portegies Zwart 2000).  It is not
obvious whether all young SSCs have `normal' MFs.  For example, the
best studied Galactic example, NGC~3603, has a normal MF extending
down to $<$1~\Msun according to Eisenhauer et al. (1998).  The compact
young Galactic centre clusters, the Arches and Quintuplet, appear to
have remarkably flat MFs for M$>$10~\Msun (Figer et al. 1999),
although they have probably undergone rapid mass segregation (Kim,
Morris \& Lee 1999) and could even represent the dense cores of
dissolved clusters (Gerhard 2001).

In the LMC SSC R136, Massey and Hunter (1998) demonstrated a normal
mass function for stars with $M \geq 2.8$\,M$_\odot$ while Sirianni et
al. (2000) find that the mass function flattens below 2\,M$_\odot$;
the R136 MF could be slightly `top-heavy' compared to that for LMC
field stars.  Similarly, Sternberg (1998) suggests a deficiency of low
mass stars in NGC~1705-1.  Spectra of SSCs in the peculiar galaxy
NGC~1275 taken with the Keck I telescope show unusually strong
H-Balmer lines; Brodie et al. (1998) interpret these as evidence for
flat MFs in clusters that are about 350~Myr old.

More generally there has been much debate over the slope of the IMF
in starburst galaxies, and particularly, in M82. Rieke et al. (1980)
first suggested that the M82 starburst has a deficit of stars below
3\,M$_\odot$ on the basis of its high $K$ band luminosity and
relatively small dynamical mass. Rieke et al. (1993) used up-dated
models and observational constraints from McLeod et al. (1993) to
demonstrate further that the M82 starburst has an MF biased towards
more massive stars.  Doane \& Mathews (1993) also find that the M82
MF must be top-heavy to explain both the high supernova rate and low
dynamical mass. On the other hand, Satyapal et al.  (1995, 1997)
derived a smaller value for the extinction-corrected $K$ band luminosity
and found that the MF could be consistent with a Salpeter slope.

In a previous paper (Gallagher \& Smith 1999; hereafter Paper I), we
presented a spectroscopic study of two M82 SSCs, denoted F and L by
O'Connell \& Mangano (1978). M82-F is one of the most luminous SSCs
known (O'Connell et al. 1995) and is located 440\,pc south-west of the
nucleus. We found that its radial velocity indicated that it could be
buried deep within M82 and fortuitously viewed through a hole in the
dust layer. Conversely, the nearby cluster M82-L is highly reddened
but its radial velocity indicates that it lies near the middle of the
M82 disc. Our blue spectrum of M82-F shows mid-B type spectral
features and the red spectra of M82-F and M82-L have a strong Ca\,{\sc
II} triplet and numerous F- and G-type absorption lines. We compared
the blue spectrum of M82-F with theoretical model cluster spectra
using the {\sc PEGASE} spectral synthesis code (Fioc \&
Rocca-Volmerange 1997), and derived an age of $60\pm20$\,Myr.  For
M82-L, we find that the similarities in the strength of the Ca\,{\sc
II} triplet and overall spectral appearance with M82-F suggest a
similar age.

The location away from the central starburst and compactness of M82-F
make it an ideal candidate in M82 for determining its dynamical mass
by measuring its size and internal velocity dispersion.  It is an
important addition to SSCs with dynamical masses because it is older
and probably more metal-rich than the two SSC dynamical mass
measurements by Ho \& Filippenko (1996a,b), but has the disadvantage
of being relatively faint, $V=16$.  In this paper we present the
results of such a study based on high dispersion echelle spectroscopy
obtained at the 4.2-m William Herschel Telescope (WHT).  We compare
the dynamical mass of M82-F with those determined for other young SSCs
by Ho and Filippenko (1996a,b), using up-dated parameters.  In
particular, we discuss its luminosity-to-mass ratio in the context of
IMF models that lack lower mass stars. The form of the IMF in turn
affects the ability of the cluster to survive as a bound system and
thus its status as a young globular star cluster.  We also discuss the
dynamics of neutral and ionized gas in the line of sight to M82-F
through the analysis of the Na\,{\sc I} interstellar lines and various
H\,{\sc II} region emission lines.

\section{Observations and Data Reduction} 
\subsection{Images}\label{images}
\begin{figure*}
\epsfysize=20cm
\epsfbox{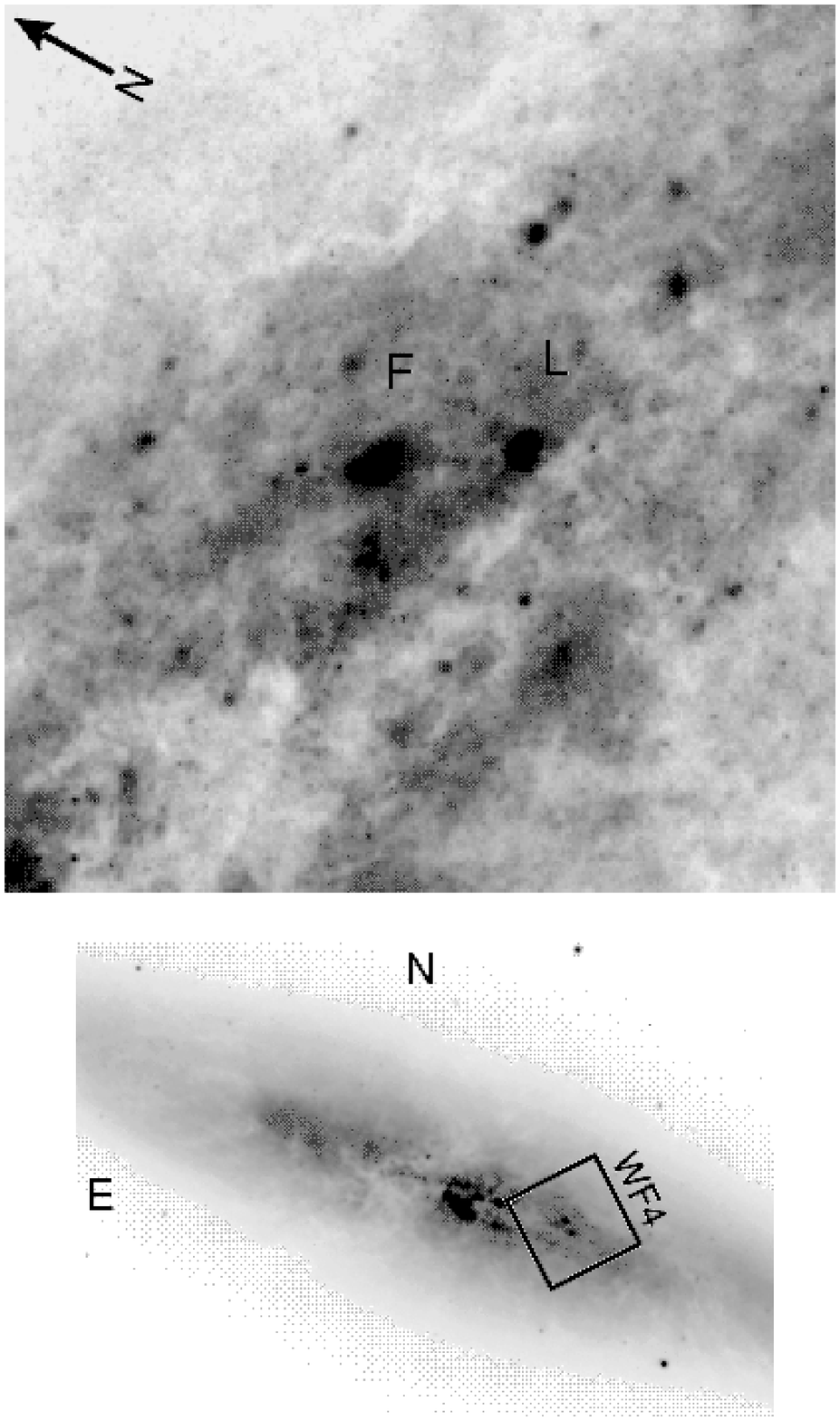}
\caption{
These $I$-band images show the locations of clusters F and L in M82.  The
lower picture was taken with a CCD camera on the WIYN 3.5-m
telescope in $\approx1$~arcsec seeing. The location of the 25$\times$25
arcsec section of the WFPC2 WF4 CCD containing the
clusters is marked.  The upper image is from the WFPC2 F814W observations,
and displays the complex small scale structure of star clusters and dust
lanes in the region of M82 where clusters F and L are located.}
\label{hst_wf}
\end{figure*}
In Fig.~\ref{hst_wf}, the locations of clusters F and L within M82 are
shown using $I$-band images taken with the WIYN~3.5-m telescope and
the Wide Field Planetary Camera (WFPC2) onboard {\it HST}. The WFPC2
image was obtained from the {\it HST} archive and is an average of two
200\,s exposures in the F814W filter.  There are also $V$-band (two
350~s exposures in the F555W filter) and $B$-band (two 400~s exposures
in the F439W filter) images available. These images were taken as part
of the GO 7446 program (R. W. O'Connell, P.I.) to investigate star
clusters and other features of the visible surface of the M82
starburst (see de Grijs et al. 2000, 2001). They were not designed for
measurements of bright star clusters such as M82-F, and this
introduced some complications.
\begin{figure*}
\epsfysize=8cm
\epsfbox{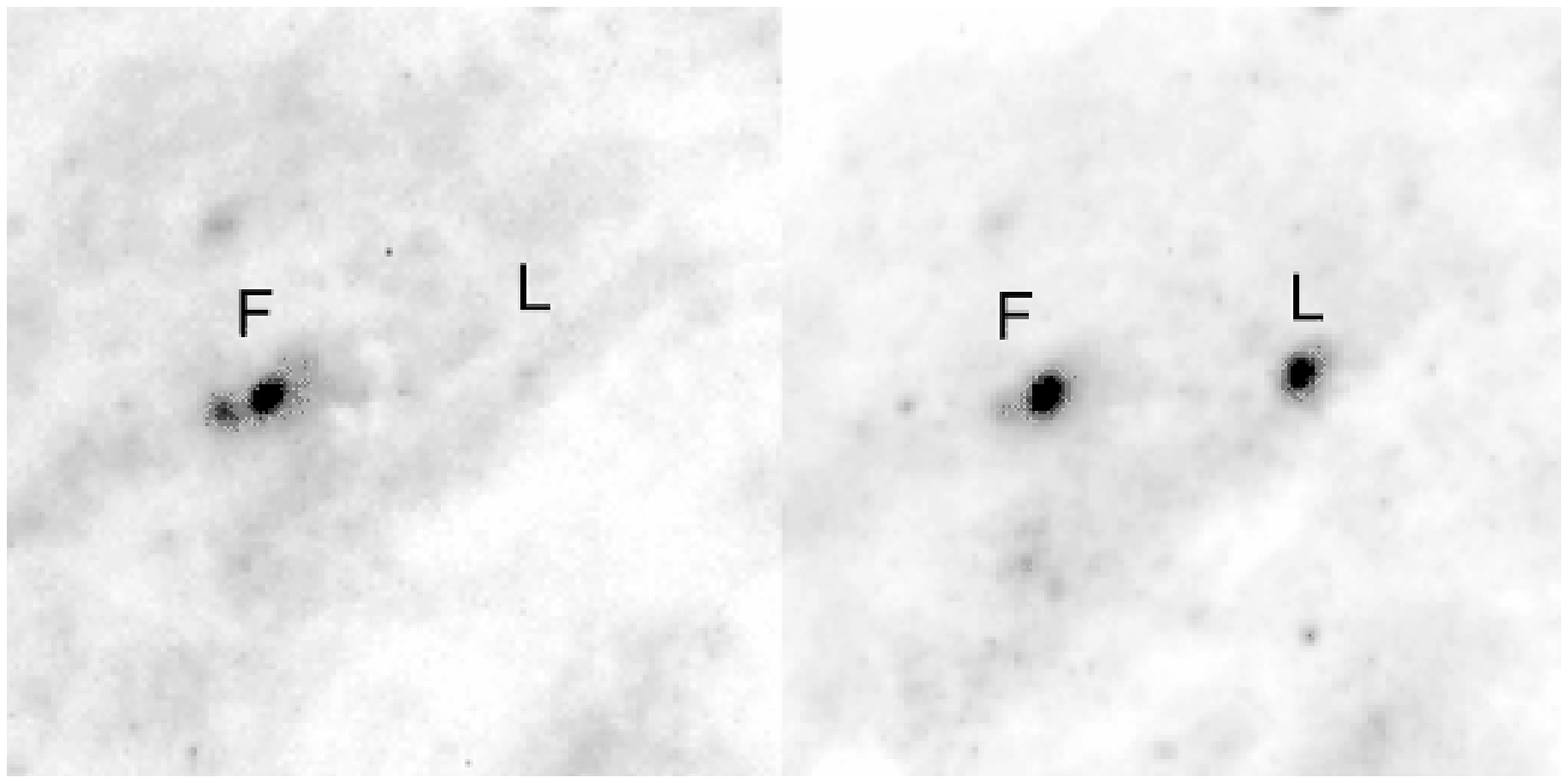}  
\caption{ Blue (left panel, F439W filter) and red (right panel, F814W
filter) WFPC2 {\it HST} observations of an 8~arcsec field around M82
clusters F and L are shown. M82-L is so heavily reddened as to be
invisible in the blue, while the companion cluster located to the left
and slightly below M82-F is most pronounced in the blue.}
\label{hst_F_BI}
\end{figure*}

Images of M82-F and M82-L are both on the WFPC2 WF4 CCD, which has an
image scale of 0.1~arcsec pixel$^{-1}$. We worked from combined pairs
of images in each filter, and used standard {\sc IRAF}\footnote{The
Image Reduction and Analysis Facility (IRAF) is distributed by the
National Optical Astronomy Observatories which is operated by the
Association of Universities for Research in Astronomy, Inc. under
cooperative agreement with the National Science Foundation. STSDAS is
the Space Telescope Science Data Analysis System; its tasks are
complementary to those in IRAF.}  software to remove cosmic rays
(there were no major cosmic ray hits on or near M82-F). The F439W
short exposure images are unsaturated and give a true measure of the
M82-F brightness profile. Both the F555W and F814W images contain 9
saturated pixels (due to the WFPC2 A/D converter response) in the
M82-F cluster centre.  We corrected these by using the F439W image as
a template and replaced the saturated pixels with values scaled from
the surrounding regions of the cluster in the F555W and F814W bands.
To estimate the error, we arbitrarily increased the values of all
saturated pixels by 30 per cent, and find that this leads to an
increase of slightly more than 0.1 magnitudes in the brightness of
M82-F in both the F555W and F814W images.  The method we have used
assumes that there is no radial colour gradient present in M82-F. Any
colour changes are expected to be small since typical gradients of at
most $\sim 0.1$ mag are observed in rich LMC clusters (Elson, Fall \&
Freeman 1987). We therefore adopt an error of $\pm 0.10$ mag for the
$V$ and $I$ magnitudes of M82-F.
The F439W and F814W images of M82-F are shown in
Fig.~\ref{hst_F_BI}. M82-L is very bright in the F814W-band but
essentially absent in F439W. We also see that M82-F has a companion
star cluster at a projected distance of 0.7~arcsec along a position
angle of 50$^\circ$.

Aperture photometry was performed on M82-F and its companion.  The
region of the companion cluster was replaced with the mean background
for the M82-F photometry. A circular aperture with a 7 pixel radius
was chosen for the total magnitude measurement. The background was
measured from the centroid of the distribution of pixel brightnesses
in an annulus at 12--14 pixels. Magnitudes were transformed to the
Landolt $BVI$ system following the precepts of Holtzman et
al. (1995). We obtain $V=16.12$ and $V-I=1.51\pm0.10$
(Table~\ref{tab_par}).  Using a smaller aperture of radius 0.5~arcsec,
we obtain $V=16.31\pm0.10$; O'Connell et al. (1995) used observations
taken with the original WFPC to estimate a deconvolved $V=16.3$ within
the same radius.

Photometry of the companion star cluster was performed by estimating a
constant background which was derived by assuming that M82-F has a
symmetric structure, and then using the mean count rates on the side
opposite of the cluster to the companion. This procedure yielded for
the M82-F companion cluster $V=18.4 \pm 0.2$ and $B-V=0.9 \pm 0.3$
using a 5 pixel radius circular aperture.  Due to the faintness of
this object with respect to the galaxy background and proximity of
M82-F, we could not obtain reliable photometry from the F814W WFPC2
image.

From circular aperture photometry on M82-F, we measure a half-light
radius $r_{\rm h}$(F439W)$ = 0.22\pm0.02$~arcsec (1 pixel $=
0.1$~arcsec).  The ratio of major to minor axes was determined to be
about 1.3 using an ellipse fitting routine in the {\sc IRAF} software
package.  This gives a negligible ellipticity correction of
$<0.01$~arcsec for the effective radii of circular apertures used for
the photometry. To correct for the effects of blurring of the
point-spread-function (PSF) of the WF cameras of WFPC2, we used the
simple approach of Phillips et al. (1997) who suggest subtracting
$r_{\rm h}= 0.15$~arcsec from the observed value in quadrature.  This
then gives a projected half-light radius of $r_{\rm h}= 0.16$~arcsec
or $2.8\pm 0.3$\,pc for an adopted distance to M82 of 3.6\,Mpc.
O'Connell et al.  (1995) derived dimensions of $9 \times 5$\,pc or a
mean FWHM of 2.8\,pc from their deconvolved WFPC images.
\begin{figure*}
\epsfysize=8cm
\epsfbox[50 260 550 530]{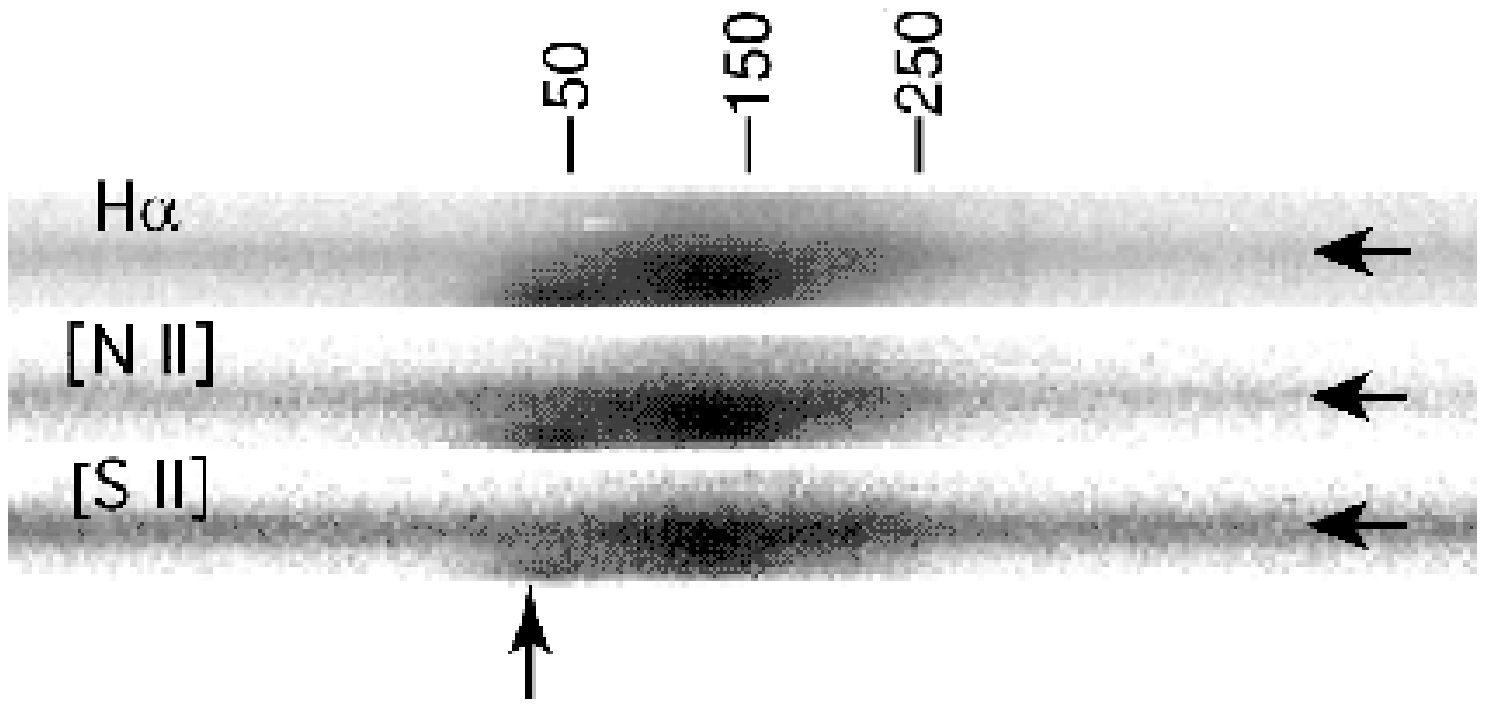}
\caption{
Greyscale images are shown for the individual orders of echelle spectra
from the first night of WHT observations. The major emission lines 
are marked, where we have chosen the stronger of the [N\,{\sc II}] 
and [S\,{\sc II}] 
doublets. A heliocentric velocity scale is plotted above the spectra, and 
the location of M82-F along the 6~arcsec slit in each echelle spectral
order and in radial velocity is marked with arrows.}
\label{ech_vel}
\end{figure*}

The SSC NGC~1705-1 is an important comparison object for M82-F.  Short
exposure, unsaturated archival WFPC2 images of NGC~1705-1 on the PC
are available in the F380W and F439W filters from {\it HST} program
GO~7506 (M. Tosi, P.I.). The guiding in these exposures is not good.
Images are elongated in both exposures, with the problem being more
serious in the F439W than F380W image. These images show an extended
halo of luminous stars around NGC~1705-1 and a bright object nearby
that is identified as the second brightest cluster in the galaxy by
O'Connell et al. (1994).

The half-light radius of NGC~1705-1 is uncertain, with O'Connell et
al.  (1994) estimating 0.14 arcsec and Meurer et al. (1995) re-deriving
0.04 arcsec from the same WFPC PC data. We therefore measured $r_{\rm
h}$ from the F380W archival WFPC2 image which appears to have the best
image quality.  Our technique is the same as that used for M82-F. A
PSF correction was derived from measurements of stars surrounding the
cluster to be 1.8 WFPC2/PC pixels, or 0.082~arcsec. The measured
projected half-light radius for the central star cluster measured from
multiple aperture photometry is 2.25 pixels, and thus $r_{\rm
h}=0.062\pm0.01$~arcsec. O'Connell et al. (1994) measured a distance
to NGC 1705 of $5.0\pm2.0$~Mpc; this has been revised recently by
Tosi et al. (2001, in prep.) to $5.3\pm0.8$~Mpc from the tip
of the red giant branch. We adopt this new distance and find that
$r_{\rm h}=1.6\pm0.4$~pc.

We also re-measured the total magnitude of {NGC~1705-1} from aperture
photometry of the 140~s exposure F439W image. The photometry is
complicated by the presence of the nearby bright cluster (at a
projected distance of 0.9 arcsec) and the extended halo.  Aperture
photometry with a radius of 25 pixels ($=1.2$ arcsec), a reasonable
estimate of the maximum cluster size, gives $m_{{\rm F439W}}=14.85$
with small aperture photometry of the companion cluster subtracted.
This SSC has $B-V \approx0$, in which case $m_{{\rm F439W}} \approx B
=14.9\pm0.1$, where the error range includes a provision for
transformation problems (see Holtzman et al. 1995).  Aperture
photometry to 50 pixels with the companion cluster subtracted gives
$B=14.75\pm0.1$. This value agrees well with the value of $V=14.7$
reported by O'Connell et al. (1994) for {NGC~1705-1} for a 3.4~arcsec
radius aperture. We adopt $B=V=14.9\pm0.1$ for NGC 1705-1.

\subsection{Spectra}
M82-F was observed with the Utrecht echelle spectrograph (UES) at the
Nasmyth focus of the WHT on 1999 February 26 and 27. The 31.6 grooves
mm$^{-1}$ echelle grating was used with a $2048 \times 2048$ SITe CCD
to record the wavelength range 5760--9140 \AA\ in 36 orders.  The
orders overlap by 34--55\,\AA\ since the size of the detector exceeds
the free spectral range of this grating (59--143\,\AA).  M82-F was
observed with a slit length of 6 arcsec (corresponding to 18 spatial
pixels) to ensure no overlap between adjacent orders while still
providing a reasonable sample of the background. The slit was kept at a
position angle of $139^\circ$ throughout the observations. A slit width
of 1.3 arcsec was used which projected to 2.3 detector pixels on the
detector (as measured from ThAr comparison arc lines), giving a
resolving power of 45\,000. The total integration time over the two
nights was 23\,200\,s in seeing conditions of 0.8--1\,arcsec. In
addition, two atmospheric standards and eight template stars were
observed ranging in spectral type from A7\,III to M2\,Iab.

The echelle data were reduced  using the {\sc FIGARO} software package
(Shortridge et al. 1997). Each  frame was cleaned of cosmic rays, bias
corrected  and divided  by  a normalized  flat  field. The  individual
orders were next extracted for the object and two adjacent
background regions (sky $+$ galaxy) and  wavelength
calibrated.   The  orders were  then  background-subtracted, binned
to  4\,km\,s$^{-1}$ wide  intervals, and co-added.  Finally the orders
were merged  in the  overlap regions  and normalized  to unity using
spline  fits. The  final spectrum of  M82-F has a  resolution of
8\,km\,s$^{-1}$  and  a  signal-to-noise  per  pixel  of  15--25.  The
template spectra were reduced in exactly the same manner.

\section{Interstellar Medium}
\subsection{Ionized Gas}\label{ionized}
To investigate the dynamics of the ionized gas in the line of sight to
M82-F, we simply took the median of the echellograms obtained on the
first night since this night had the best seeing conditions. The
resulting image is shown in Fig.~\ref{ech_vel} in the region of the
H$\alpha$, [N\,{\sc II}] and [S\,{\sc II}] emission lines.  All these
emission lines show an identical dynamical structure.  Examining the
line profiles along the 6 arcsec slit ($=18$ pixels), we find that
there are at least three components present at $+50$, $+120$ and
$+180$ km\,s$^{-1}$ with an overall FWHM of $\sim 120$
km\,s$^{-1}$. The peak of the CO rotation curve lies between the 120
km\,s$^{-1}$ and 50 km\,s$^{-1}$ emission components (Shen \& Lo
1995).  The heliocentric radial velocity of M82-F is measured to be
$+24$ km\,s$^{-1}$ from the Ca\,{\sc II} triplet lines, in good
agreement with the value of $+35$ km\,s$^{-1}$ obtained in Paper I.

The first two velocity components are resolved in the first three
spatial pixels but appear to merge to produce two components at $+124$
and $+180$ km\,s$^{-1}$ at the position of M82-F on the slit
(Fig.~\ref{ech_vel}).  The strongest component at $+124$ km\,s$^{-1}$
then weakens and the last three spatial pixels show an unresolved broad
(FWHM $=135$ km\,s$^{-1}$) emission feature at $+137$ km\,s$^{-1}$. In
Paper I we found that the same emission lines are split into two
components at $+114$ and $+181$ km\,s$^{-1}$ with the lower velocity
component occurring along the sight line between M82-F and M82-L. The
positions angles used in the two sets of observations are quite
different: $139^\circ$ for the UES and $20^\circ$ for the observations
in Paper I. In this region of M82 the CO emission extends over
$\approx 150$ km\,s$^{-1}$ and shows a pronounced extension towards
low radial velocities that also is seen in radio H-recombination lines
(e.g. Seaquist et al. 1996).  We conclude that no association exists
between ionized gas and M82-F; this is not surprising given the age
and high orbital velocity of the cluster (see Sect.~\ref{survival}).
\begin{figure}
\epsfxsize=8.3cm 
\epsfbox[114 187 413 582]{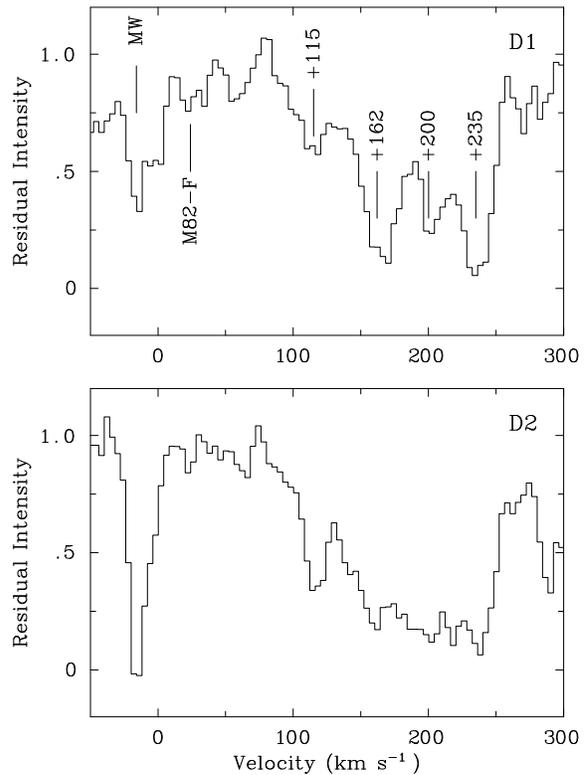}
\caption{Portion of the WHT$+$UES spectrum of M82-F showing the region
of the interstellar Na\,{\sc I} D lines. The velocity scale is heliocentric
and the spectra are normalized. The velocities of the resolved components
are indicated. The strong narrow feature at $-16$ km\,s$^{-1}$ is due to
interstellar absorption in the Milky Way. The radial velocity of M82-F
is also indicated.}
\label{NaD_plot}
\end{figure}

\subsection{Neutral Sodium Absorption}
Figure~\ref{NaD_plot} shows the Na\,{\sc I} D$_1$ and D$_2$ absorption
line profiles observed towards M82-F. The presence of multiple sharp
components in the D$_1$ line demonstrates that this feature is due to
intervening interstellar gas rather than stellar photospheres as
assumed in some previous studies (e.g. Sait\={o} et al. 1984; G\"otz
et al. 1990).  The individual components can be seen more clearly in
the D$_1$ line, indicating that optical depth variations are occurring
between the D$_1$ and D$_2$ lines; these features are formed in low
column density neutral gas in M82. The heliocentric centroid velocity
of about 190~km~s$^{-1}$ agrees with the Na\,{\sc I} absorption radial
velocities in Paper I.
\begin{figure*}
\epsfxsize=17cm 
\epsfbox[48 244 518 556]{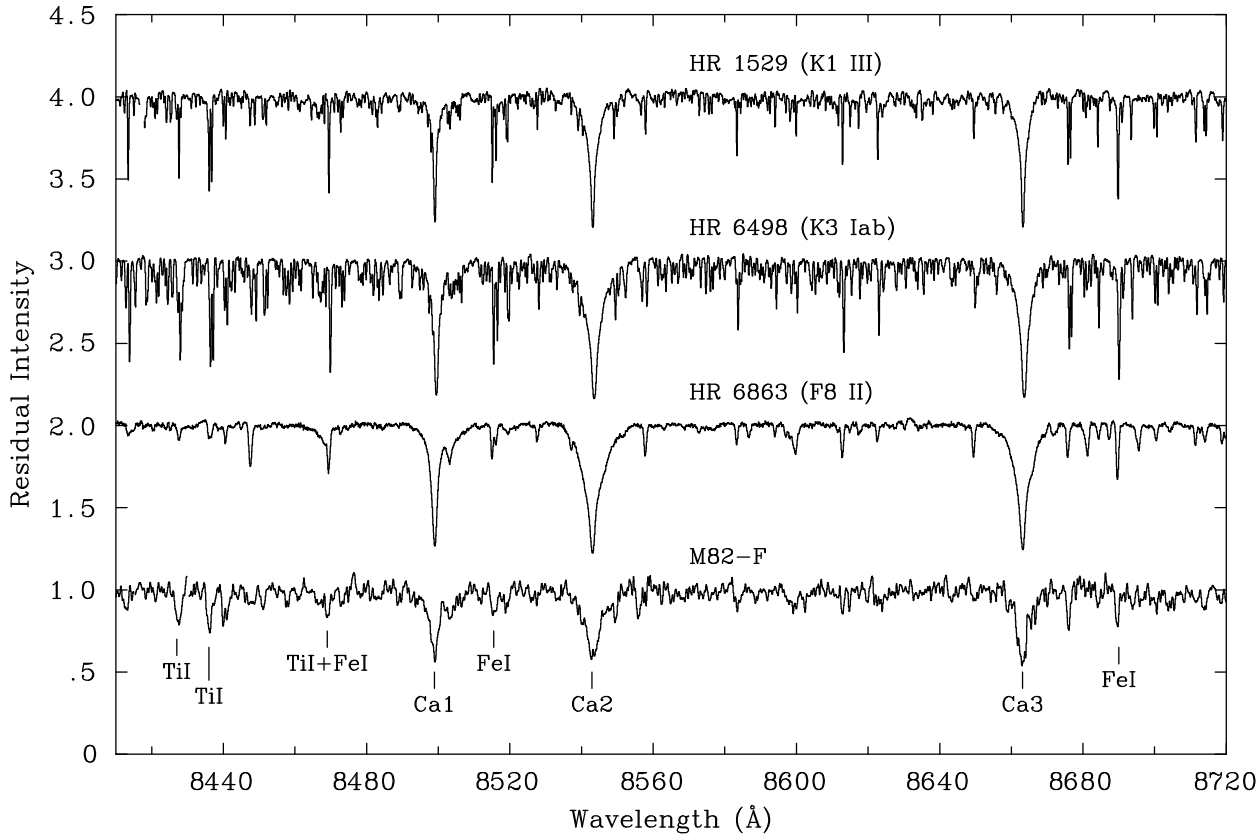}
\caption{WHT$+$UES spectra covering the Ca\,{\sc I}I triplet region at
a resolution of 8 km\,s$^{-1}$ for M82-F and the three template stars
which show the closest spectral match.  The spectra have been
normalized and shifted in velocity space to match the radial velocity
of M82-F, and the M82-F spectrum has been lightly smoothed for
presentation purposes. Absorption features due to the Ca\,{\sc I}I
triplet and Fe\,{\sc I} and Ti\,{\sc I} are indicated.}
\label{CaT}
\end{figure*}

With the resolution of the echelle, we see that these lines extend
from approximately 115 to 240~km~s$^{-1}$ at a location in M82 where
the rotation peak is blueshifted to $\sim$ 80~km~s$^{-1}$ (see
Sect.~\ref{ionized}).  Adopting a systemic velocity of
$200\pm10$~km~s$^{-1}$ and a rotational velocity amplitude of
$\approx100$~km~s$^{-1}$ for M82 (Yun, Ho \& Lo 1993; Achtermann \&
Lacy 1995), we see that absorbing gas in circular orbits within
the disc must lie in the approximate velocity range of 100~km~s$^{-1}
< V_c <$200~km~s$^{-1}$ i.e. between the rotation curve peak and zero
velocity relative to the M82 system. Figure 4 shows this is where the
bulk of the Na\,{\sc I} absorption is located, although lines
redshifted beyond the systemic velocity are forbidden for circular
rotation. This gas could be associated with infalling tidal
debris or gas outflow from M82, although the H\,{\sc I} map of Yun et
al. (1993) suggests that most of the gas near M82-F has a velocity of
less than 150~km~s$^{-1}$.

On the other hand, the CO 1-0 maps of Shen \& Lo (1995) and Sofue et
al. (1992) show a large redward velocity extent to the south-west of
the nucleus that becomes more pronounced above the galaxy's
midplane. Since M82 is seen nearly edge on, the location of M82-F with
respect to the mid-plane is uncertain, and we could be sampling some
of this gas.  Shen \& Lo (1995) propose that the high velocity
dispersion molecular emission originates in spiral arms where previous
starburst activity disrupted the interstellar medium. High angular
resolution maps of gas kinematics in the outer disc of M82 would be
very useful in disentangling the multiple interactions in this
peculiar system.

\section{Parameters of M82-F} 
The wavelength region of the UES observations (5760--9140\,\AA) was
chosen because, as discussed by Ho \& Filippenko (1996a,b), and
demonstrated by our intermediate dispersion spectra of M82-F (Paper
I), the spectral region longward of 5000 \AA\ is dominated by the
light of cool evolved stars.  Thus by cross-correlating the cluster
spectrum with a suitable template spectrum, it is possible to recover
the line-of-sight velocity dispersion of the cluster, and hence derive
its dynamical mass by application of the virial theorem.

\subsection{The Velocity Dispersion}
\begin{figure}
\epsfxsize=8.3cm
\epsfbox[59 72 445 710]{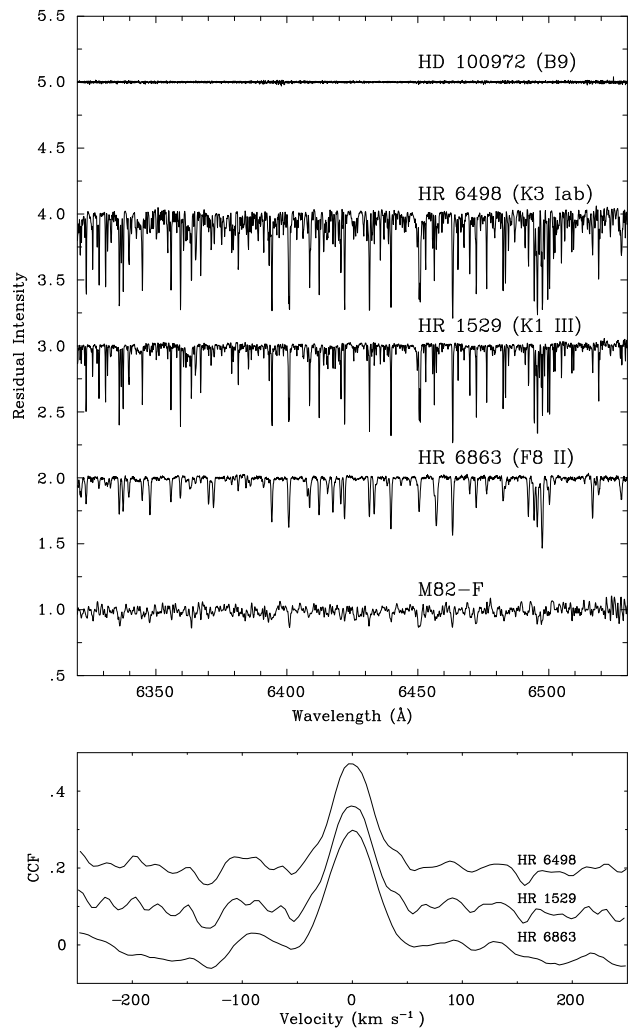}
\caption{{\it Top panel:\/} Comparison of the normalized spectra of
M82-F and the three template stars for one of the wavelength regions
used for the cross-correlation.  The spectrum of the B9 atmospheric
standard is also shown to demonstrate the absence of telluric
features.  Again the spectra have been shifted to match the velocity
of the cluster, and that of M82-F is lightly smoothed for presentation
purposes.  {\it Bottom Panel:\/} The cross-correlation functions (CCFs)
between the cluster and the three template stars. The velocity
dispersion of M82-F can be derived from the FWHM of the CCF.}
\label{ccf}
\end{figure}

The spectrum of M82-F was compared with the spectra of the eight
template stars to find the best match in spectral type. In
Fig.~\ref{CaT}, we show the region containing the Ca\,{\sc II} triplet
for M82-F and the closest three template stars: HR 6863 (F8\,II), HR
1529 (K1\,III) and HR 6498 (K3\,Iab).  The metal lines in this
wavelength region are mainly due to Fe\,{\sc I} and Ti\,{\sc
I}. Comparison of the ratios of the strengths of these lines to the
Ca\,{\sc II} triplet lines shows that the closest spectral match is to
the two K stars.  We have used these three stars as templates and
cross-correlated their spectra with that of M82-F. We chose four
spectral regions which are free of telluric absorption lines:
6010--6275 \AA, 6320--6530 \AA, 7340--7590 \AA, and 7705--8132 \AA\
for the analysis.  All these regions gave a strong cross-correlation
signal; an example is shown in Fig.~\ref{ccf} where we also show the
spectrum of an atmospheric standard.  We ignored the region containing
the Ca\,{\sc II} triplet because these lines are saturated in the
template spectra and broad Paschen absorption lines from early-type
stars are present in the M82-F spectrum.

The resulting cross-correlation peaks were fitted with Gaussian
profiles using least-squares to obtain the FWHMs. The relationship
between the FWHM and the velocity dispersion was empirically
calibrated by broadening the template spectra with Gaussians of
different velocity dispersions and cross-correlating with the original
spectra. We find that relative to HR 6863 (F8\,II), HR 1529 (K1 III)
and HR 6498 (K3 Iab), the velocity dispersion of M82-F is $13.4\pm0.7$,
$15.4\pm1.6$ and $15.3\pm1.5$ km\,s$^{-1}$ respectively, where the
errors represent the $1\sigma$ dispersion on the mean.  The values
derived using the F and K template spectra are different because the
intrinsic photospheric absorption line widths are smaller in the two K
stars. It is surprising to find that the K3 Iab star
in our sample has the same line widths as the K1 giant, suggesting the
K3 star may be wrongly classified in terms of its luminosity.
The velocity dispersions derived from cross-correlating the two
K stars (HR 1529 and 6498) with the F8 bright giant are $8.8\pm0.6$
and $8.7\pm0.3$ km\,s$^{-1}$ respectively, and are entirely consistent
with the differences in the cluster velocity dispersions found from the
various template stars.

To choose the appropriate velocity dispersion value for M82-F, we need
to determine which template star provides the best spectral/luminosity
match and consider the effects of macroturbulent broadening in the
atmospheres of cool evolved stars.  In bright giants macroturbulence
is the dominant factor in setting weaker stellar absorption line
profile widths, and this depends primarily on luminosity class for
cooler stars. As discussed above, the closest spectral match is to
the two K stars. Inspection of the Starburst99 code (Leitherer et al.
1999) shows that K\,II stars dominate the UES wavelength region at an
age of 60 Myr (Paper I). We thus choose the F8\,II template star since
Gray \& Toner (1986) show that the macroturbulent broadening in F5--K4
luminosity class II bright giants is very similar with a mean value
of $7.2\pm1.5$ km\,s$^{-1}$, while the K\,III stars have sharper lines,
in agreement with our results.

We therefore adopt a velocity dispersion $\sigma$ of $13.4\pm0.7$
km\,s$^{-1}$ for M82-F.  {\it M82-F has a
sufficiently large stellar velocity dispersion to be a proto-globular cluster.}

\subsection{The Dynamical Mass}
The mass of M82-F can be obtained from the virial theorem since the
average stellar crossing time of $\approx 4 \times 10^5$ yr is much
less than the cluster age, and thus the cluster should be near virial
equilibrium.  The virial mass $M$ is given by
\[ M = 7.5\,{{\sigma^2 r_{\rm m}}\over {G}}\]
where it is assumed that the cluster is spherical with an isotropic
velocity distribution such that $\sigma$ represents the
one-dimensional rms velocity dispersion along the line-of-sight
(Spitzer 1987).  With the radius expressed in terms of the half-mass
radius $r_{\rm m}$ (rather than the gravitational radius), the
constant of proportionality is 7.5 (see e.g. the review of Gerhard
2000). Further assuming that the half-light radius derived from the
WFPC2 image (Sect.~\ref{images}) represents the projected half-mass
radius and taking $r_{\rm m} = 4/3\,r_{\rm h}$ (Spitzer 1987), we
derive a mass of $1.2\pm0.1 \times 10^6$\,M$_\odot$, and a mass
density within the half-mass radius of
$6.4\pm0.9~\times~10^3$\,M$_\odot$\,pc$^{-3}$.  M82-F is clearly a
very massive young cluster that might become a globular cluster.

\subsection{The Luminosity-to-Mass Ratio}
\begin{table}
\caption {Summary of derived parameters for M82-F}
\begin{tabular}{ll}
\\
\hline
Parameter & Value\\
\hline
Age & $60\pm20$ Myr \\
Half-light radius $r_{\rm h}$& $2.8\pm0.3$ pc \\
$V$  & $16.12\pm0.10$ \\
$(B-V)$  & $1.22\pm0.10$ \\
$(V-I_{\rm c})$  & $1.54\pm0.10$ \\
Radial velocity $V_{\rm r}$ & $+24.3\pm1.7$ km\,s$^{-1}$ \\
Velocity dispersion $\sigma$ & $13.4\pm0.7$ km\,s$^{-1}$ \\
Mass $M$ & $1.2\pm0.1 \times 10^6$ M$_\odot$ \\
Density $\rho$ & $6.4\pm0.9 \times 10^3$\,M$_\odot$\,pc$^{-3}$ \\ 
$E(B-V)$ &$0.9\pm0.1$ \\
$M_V$ & $-14.5\pm0.3$ \\
Log $L_V/L_\odot$ & $7.73\pm0.12$ \\
$(L_V/M)_\odot$ & $45\pm13$ \\
\hline
\\
\end{tabular}
\label{tab_par}
\end{table}
One of the most important parameters to derive when considering
whether M82-F will survive to become an old globular cluster is the
luminosity-to-mass ratio $(L_V/M)_\odot$ (see discussion in
Sect.~\ref{intro}).  This tells us if a cluster has sufficient mass in
the form of lower luminosity, long-lived stars to avoid  possible
dynamical disruption.  To determine this ratio we require the absolute
magnitude and thus the reddening towards M82-F. O'Connell \& Mangano
(1978) estimated $E(B-V) \approx 1.0$. In Paper I, we estimated a
reddening of $\sim1.5$ by comparing the theoretical flux distribution
of a 60\,Myr model cluster spectrum to the blue spectrum of M82-F. We
found that a standard Galactic extinction model did not correct the
data in the near-UV. We now realise that this disparity is
partially due to contamination from the close companion at the bluest
wavelengths since this cluster was in the slit.

We are now able to obtain an accurate value for the reddening towards
M82-F by comparing the WFPC2 $(B-V)$ and $(V-I)$ observed colours with
the colours predicted for a 60 Myr old cluster.  From the spectral
synthesis models, Starburst99 (Leitherer et al. 1999) and {\sc PEGASE} (Fioc
\& Rocca-Volmerange 1997), we derive $E(B-V)=0.86$ and 0.82 using the
observed $(V-I)$ colour given in Table~\ref{tab_par}, and correcting
for the different photometric systems used in the two synthesis
models. From the observed $(B-V)$ colour, we obtain $E(B-V)=1.11$ and
thus adopt a reddening towards M82-F of $0.9\pm0.1$ where the error
reflects the range in the reddening derived from the different
photometric indices.
\begin{figure}
\epsfxsize=8.3cm 
\epsfbox[92 229 449 576]{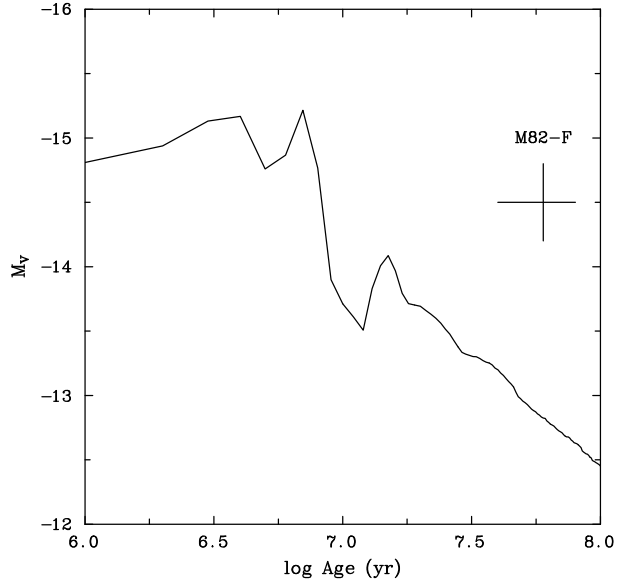}
\caption{The absolute magnitude $M_V$ plotted as a function of age for
a cluster of the same mass as M82-F at 60~Myr for solar metallicity, a
Kroupa (2001) IMF ($\alpha=1.3$ for masses in the range 
$0.1\le M <0.5$~$M_\odot$ and $\alpha=2.3$ for $M \ge 0.5~$M$_\odot$),
lower and upper mass limits of 0.1 and
100~M$_\odot$, calculated using the Starburst99 code (Leitherer et al.
1999). The derived $M_V$ and age for M82-F are plotted. It is clearly
too bright for its age by $\sim 1.7$ mag, indicating that the IMF as
shown is not appropriate for this cluster.}
\label{mv}
\end{figure}

In Table~\ref{tab_par}, we list all the parameters derived for M82-F.
Using these values, we obtain $M_V=-14.5\pm0.3$, and a
luminosity-to-mass ratio $(L_V/M)_\odot = 45\pm13$. In Fig.~\ref{mv},
the predicted variation of $M_V$ with age is shown for a cluster of
the same mass as M82-F at 60~Myr with a standard IMF where the slope
is flatter below 0.5\,M$_\odot$. We have adopted the parameterisation
of Kroupa (2001): $\alpha=1.3$ for masses in the range 
$0.1\le M <0.5$~$M_\odot$ and $\alpha=2.3$ for $M \ge 0.5~$M$_\odot$.
We used the
Starburst99 population synthesis code (Leitherer et al. 1999) with
lower and upper mass limits equivalent to 0.1 and 100~M$_\odot$. 
Mass-loss due to stellar winds and supernovae are
accounted for using the mass-loss rates of the Geneva models  and
assuming that stars more massive than 8\,$M_\odot$ explode as
supernovae and leave 1.4\,$M_\odot$ remnants.

It is clear that M82-F is too bright for its age by $\approx
1.7$~mag (a factor of 5).  A standard IMF would give an acceptable
fit to the measured $(L_V/M)_\odot$ only if M82-F was near the peak
of its optical luminosity, with an age of $\sim 8$~Myr. {\it We thus
find that our derived parameters for M82-F strongly suggest an
abnormal MF.}  
We now review the likely sources of error and their impacts on our
derived parameters for M82-F before discussing this result. The
considerable discrepancy shown in Fig.~\ref{mv} indicates that we are
looking for large factors in the uncertainties.

\section{Uncertainties}
\subsection{Age}
The age of M82-F is probably the most critical parameter since if it
substantially younger than 60 Myr old, then this will reduce the
discrepancy we have found. We have therefore carried out a careful
re-analysis of the age of M82-F as follows.  In Paper I, we determined
an age for M82-F by comparing the observed blue spectrum from
3510--5000\,\AA\ with model cluster spectra generated from the {\sc PEGASE}
spectral synthesis code (Fioc \& Rocca-Volmerange 1997). We used the
Jacoby et al. (1984) spectral library, the `Geneva' stellar
evolutionary tracks and a Salpeter IMF.  By matching the depth of the
Balmer absorption lines and the Balmer jump, an age of $60\pm20$ Myr
was determined. We noted that the spectral library of Jacoby et
al. had lower resolution and often inferior signal-to-noise to the
WHT$+$ISIS spectrum of cluster F. Since Paper I, a new synthetic
library has become available for the H Balmer and He\,{\sc I} absorption
lines with a sampling of 0.3\,\AA\ (Gonz\' alez Delgado \& Leitherer
1999). Using this new dataset we have re-visited the age of M82-F by
performing a detailed line profile comparison for the H Balmer series
and the He\,{\sc I} lines.
\begin{figure}
\epsfxsize=8.3cm
\epsfbox[100 158 454 606]{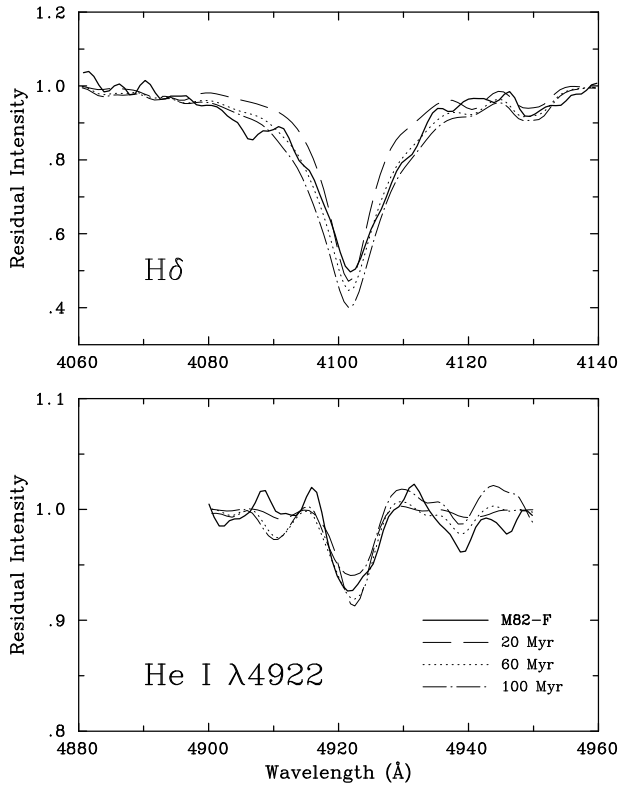}
\caption{The observed H$\delta$ and He\,{\sc I} $\lambda4922$ line
profiles for M82-F compared with synthetic profiles from the
population synthesis models of Gonz\' alez Delgado et al. (1999) for
ages of 20, 60 and 100~Myr. The M82-F data were obtained with the
WHT$+$ISIS and were presented in Paper I.  The synthetic data have
been binned to match the 1.6~\AA\ resolution of the M82-F data and all
spectra have been smoothed using $\sigma=1.0$~\AA. An age of
$60\pm20$~Myr is obtained from the best fits to the wings of the
H$\delta$ line profile (the core is affected by nebular emission) and
the absorption depth of He\,{\sc I} $\lambda4922$.  }
\label{age}
\end{figure}

We used the output of the evolutionary synthesis code Starburst99
(Leitherer et al. 1999) for an instantaneous burst and a Salpeter IMF
with upper and lower mass cutoffs of 1 and 80\,M$_\odot$ at solar
metallicity. Gonz\' alez Delgado, Leitherer \& Heckman (1999) find
that the strengths and widths of the Balmer lines increase with age
and that He\,{\sc I} absorption reaches a maximum at 10--40\,Myr and
is absent after an age of 100\,Myr. They also investigate the
sensitivity of the Balmer lines to changes in the IMF parameters and
find that a flatter IMF produces weaker Balmer features only for the
first 4 Myr because of the greater proportion of O stars on the main
sequence.

To compare observed and synthetic profiles, the M82-F spectrum was
smoothed ($\sigma=1.0$\,\AA) and pseudo-continua were fitted to the
H$\beta$--H13 and He\,{\sc I} profiles using the windows defined in Gonz\'alez 
Delgado \& Leitherer (1999). The synthetic spectra were treated in
the same way and smoothed and binned to match the resolution of the
M82-F spectrum. The observed and synthetic line profiles for H$\delta$
and He\,{\sc I} $\lambda4922$ are shown in Fig.~\ref{age} for ages of 20, 60
and 100 Myr. Since the cores of the lower members of the Balmer lines
are affected by nebular emission, we use fits to the line profile
wings as the main age discriminator. It can be seen that the 20 Myr H
line profiles are too narrow and that He\,{\sc I} $\lambda4922$ is too
strong. Conversely, at 100 Myr, the H lines are too broad and He\,{\sc I} is
too weak. We thus confirm the age determined in Paper I of $60\pm20$
Myr.

Most importantly, M82-F cannot be as young as the age of $\sim
8$~Myr required for agreement between the predicted and observed
$M_V$ shown in Fig.~\ref{mv}.  At this age, RSGs will dominate but no
molecular bands are detected (Paper I; cf. Bica, Santos \& Alloin
1990; Schiavon, Barbuy \& Bruzual 2000). The predicted $(V-I)$ colour
at 8~Myr gives a low $E(B-V) = 0.4$ in contradiction to the
$(B-V)$ colour, indicating that M82-F cannot have the colours of a
starburst dominated by RSGs. If M82-F was as young as 4~Myr then we
would expect to see Wolf-Rayet features in the spectra but these are
clearly absent (Paper I).

\subsection{Luminosity and Radius}\label{lum_r}

M82-F is observed on a complicated background, as shown in
Figs.~\ref{hst_wf} and \ref{hst_F_BI}.  We have measured this
`background' relatively far from the cluster, and so could have made
the cluster too bright. However, a 50 per cent increase in the
background only decreases the observed $V$ by 0.1~mag. Furthermore, it
seems equally likely that we could have underestimated the luminosity
by not including light outside of our 0.7~arcsec radius aperture.

The measurements of the half-light radius are reasonably robust, and
will only change slightly even if a moderate amount of luminosity has
been missed. A key assumption here is that the half-light radius is
the same as the projected half-mass radius.  This cluster appears to
be too young to have experienced significant dynamical mass
segregation since the half-mass relaxation time (Spitzer \& Hart 1971)
is $\approx 2$\,Gyr giving a mass segregation time scale of $\approx
160$\,Myr for stars with masses near the main sequence turn-off at the
age of M82-F.  Massive stars could, however, be centrally concentrated
in M82-F due to processes at birth or associated with early evolution
(see discussion in Sect.~\ref{formn}). Our procedure to correct
saturated pixels could have further exacerbated this effect if it
exists. If massive stars are extremely centrally concentrated in
M82-F, we could have underestimated $r_h$ and therefore the mass.

\subsection{Extinction} 

The range of possible interstellar obscurations remains a major source
of uncertainty. The combination of spectrophotometry in Paper I and
results presented here suggest that the colour excess is unlikely to
be $E(B-V) < 0.8$. Furthermore if we are not in the pure interstellar
extinction case, then the absorption will be larger at a fixed colour
excess (e.g. Calzetti, Kinney \& Storchi-Bergmann 1994).  The best
way to remove this as an issue would be to obtain high angular
resolution infrared $K$-band images which would minimize the effects
of interstellar obscuration. For the present it seems more likely that
we have, if anything, underestimated the amount of visual obscuration
towards M82-F.

We must also consider that we have used the predictions of stellar
synthesis codes to provide the intrinsic colour. These were based on a
Salpeter IMF with upper and lower mass cutoffs of 0.1 and
120~M$_\odot$.  For a flatter IMF with $\alpha=1.5$, the predicted
$(B-V)$ and $(V-I)$ colours differ by less than 0.1 mag at an age of
60~Myr.

\subsection{Distance}
We have assumed that M82 and M81 are at the same distance and adopt
the M81 distance of 3.6~Mpc (Freedman et al. 1994).
A new distance estimate by Sakai \& Madore (1999) places M82 about 1$\sigma$
or 0.4~Mpc further away than our adopted value. This will increase
the luminosity to mass ratio by about 10 per cent.

\subsection{Velocity Dispersion}
We adopted the velocity dispersion value derived from the
cross-correlation function of M82-F with an F8 II template star. If
instead, we had chosen a velocity dispersion value given by one of
the narrower-lined template stars, this would increase the
dispersion by a maximum amount of 15 per cent, and the mass
estimate by 30 per cent.

\subsection{Metallicity}
We have assumed that the metallicity of M82-F is solar. McLeod et
al. (1993) summarise and discuss the metallicity of M82. They conclude
that the present-day interstellar medium has solar or slightly greater
than solar metallicity.

\subsection{Mass Model}
In Sect. 4.2 we followed the approach adopted by earlier studies in using
the virial theorem applied at the half-mass radius to calculate the
dynamical mass (e.g., Ho \& Filippenko 1996a,b). However, other dynamical
models for star clusters yield different masses. For example, Dubath \&
Grillmair (1997) derived masses for globular star clusters in M31 from
integrated velocity dispersions using both King models and the virial
theorem. They find King models yield masses that are 60-80\% of those
derived from the virial theorem applied to the same globular star
clusters. That this effect is possible can be seen from the form of the
Jeans equation (e.g., Binney \& Tremaine 1987), which also could allow
our virial mass to be an underestimate in some cases.

We have also assumed that M82-F is spherical with an isotropic velocity
distibution, but the real shape is elliptical (see Fig. 2). We can
estimate the effect of an anisotropic velocity distribution on
the dynamical mass by assuming that we are viewing the cluster along one
long axis. In this case, we are likely to be overestimating the mass by
about 30 per cent since we have overestimated the internal energy. We
therefore conclude that our mass estimate could be uncertain by a factor
of (up to) $\pm$50\%  if M82-F has a dynamical structure similar to that
of globular star clusters.

In summary, we conclude that our errors appear to provide a reasonable
assessment of the uncertainties as they are currently understood. There
is no clear path that would lead to a mass increase by a factor of
several. The largest remaining ambiguity is with interstellar
obscuration, and this should be removed by future IR photometry.

\section{Parameters of SSCs NGC 1569A and NGC 1705-1} 
Stellar velocity dispersions have been measured by Ho \& Filippenko
(1996a,b) for the young SSCs NGC 1569A and NGC 1705-1. Sternberg
(1998) derived their luminosity-to-mass ratios using revised
parameters and found that NGC 1705-1 could have a top-heavy MF while
NGC 1569A has a steep MF with a slope of $\sim 2.5$. Since
this study, the structural characteristics of the two SSCs have been
further updated, and we have therefore re-derived their luminosities
and masses for comparison with M82-F, as shown in
Table~\ref{tab_comp}. The two nearest SSCs, NGC~3603 in the Galaxy and
R136a in the LMC, do not have velocity dispersion measurements and therefore
cannot be compared directly with our results for M82-F.
\begin{table}
\caption {Comparison of Properties of Super Star Clusters with Dynamical
Mass Measurements}
\begin{tabular}{llll}
\\
\hline
Cluster/ & NGC 1569A & NGC 1705-1 & M82-F \\ 
Parameter \\
\hline
\\
Age (Myr) & 4--10$^a$ &10--20$^b$ & $60\pm20$ \\
Metallicity (Z$_\odot$) & 0.20$^c$ & 0.45$^d$ & 1.0$^e$\\
M$_V$ (mag) & $-14.0\pm0.4^a$ & $-13.8\pm0.4$ & $-14.5\pm0.3$ \\
r$_h$ (pc) & $2.3\pm0.3^a$ & $1.6\pm0.4$ & $2.8\pm0.3$ \\
$\sigma$  (km\,s$^{-1}$) & $15.7\pm1.5^f$ & $11.4\pm1.5^g$ & $13.4\pm0.7$\\
Mass ($\times 10^6$M$_\odot$) & $1.3\pm0.2$ & $0.48\pm0.12$ & 
$1.2\pm0.1$ \\
$(L_V/M)_\odot$ &$26\pm11$  & $60\pm24$ & $45\pm13$ \\
\hline
\\
\noalign{
$^a$Hunter et al. (2000); 
$^b$Heckmann \& Leitherer (1997);
$^c$Kobulnicky \& Skillman (1997);
$^d$Devost, Roy \& Drissen (1997);
$^e$McLeod et al. (1993);
$^f$Ho \& Filippenko (1996a); 
$^g$Ho \& Filippenko (1996b)}
\end{tabular}
\label{tab_comp}
\end{table}

\subsection{NGC 1569A}
The star clusters in the nearby starburst galaxy NGC 1569 have been
studied most recently by Hunter et al. (2000) using new {\it HST}
WFPC2 data. We will use their parameters, adopting a distance to NGC
1569 of 2.5~Mpc (O'Connell et al. 1994). O'Connell et al. (1994) first
noted that NGC 1569A has two peaks in its light distribution from WFPC
images, suggesting it may consist of two sub-clusters A1 and A2. This
was later verified by De Marchi et al. (1997) using WFPC2 data. Gonz\'alez 
Delgado et al. (1997) suggested that the two sub-clusters have
different ages of 3 and 9~Myr on the basis of Wolf-Rayet (WR) and red
supergiant (RSG) features detected in ground-based
spectroscopy. 

Hunter et al. (2000) performed a careful analysis
of the structure of NGC 1569A and find that the two components have
similar colours, with A2 being fainter than A1 by 1.7 mag in the F555W
filter.  They suggest that there is no contradiction in the presence
of both WR and RSG stars at the same age in a low metallicity
environment, or alternatively that there may be a small age spread of
several Myr in NGC 1569A. We therefore adopt an age for this cluster
of 4--10~Myr, and assume that the velocity dispersion measurement of
Ho \& Filippenko (1996a) represents the entire cluster. We derive a
mass of $1.3\pm0.2 \times 10^6$~M$_\odot$ and $(L_V/M)_\odot = 26\pm 11$. 
The uncertainties were determined using the errors given
in Table~\ref{tab_comp}, where the error on $M_V$ is derived from the 
uncertainty in the reddening discussed by Hunter et al. (2000).

\subsection{NGC 1705-1}
NGC 1705 is a nearby dwarf galaxy which is dominated by a single super
star cluster, NGC 1705-1.  The main observational uncertainties in
previous mass derivations for NGC 1705-1 have been the half-light
radius and the distance. As discussed in Sect.~\ref{images},
measurements based on WFPC data have ranged from 0.9--3.4~pc for a
distance of $5.0\pm2.0$~Mpc (Meurer et al. 1995; O'Connell et
al. 1994). Using archival WFPC2 images, we derive $r_{\rm
h}=1.6\pm0.4$~pc, using a new distance estimate of $5.3\pm0.8$~Mpc
(Tosi et al. 2001, in prep.) With the
velocity dispersion determined by Ho \& Filippenko (1996b), we derive
a mass of $4.8\pm1.2 \times 10^5$~M$_\odot$ and $(L_V/M)_\odot
=60\pm24$, with $V=14.9\pm0.1$ (Sect.~\ref{images}).

\section{Discussion}
\subsection{The Top-Heavy MF}
In Table~\ref{tab_comp}, we list the parameters of NGC 1569A, NGC
1705-1 and M82-F. The absolute magnitude and mass of M82-F are similar
to NGC 1569A but M82-F is a factor of 3--6 times older and thus it is
exceptionally bright.  Sternberg (1998) derived $(L_V/M)_\odot = 29$
for NGC 1569A and found by comparison with population synthesis models
that this cluster has a steep MF ($\alpha \sim 2.5$) and will
probably evolve to a globular cluster-type object. Conversely, he
derived $(L_V/M)_\odot = 126$ for NGC 1705-1 and found that the MF is
either flat ($\alpha < 2$) or truncated between 1--3~M$_\odot$,
implying that this cluster will not survive. While the $(L_V/M)_\odot$
we derive for NGC 1569A is identical within the errors to the
Sternberg value, the revised value for NGC 1705-1 is a factor of $\sim
2$ lower.
\begin{figure}
\epsfxsize=8.3cm
\epsfbox[93 229 460 576]{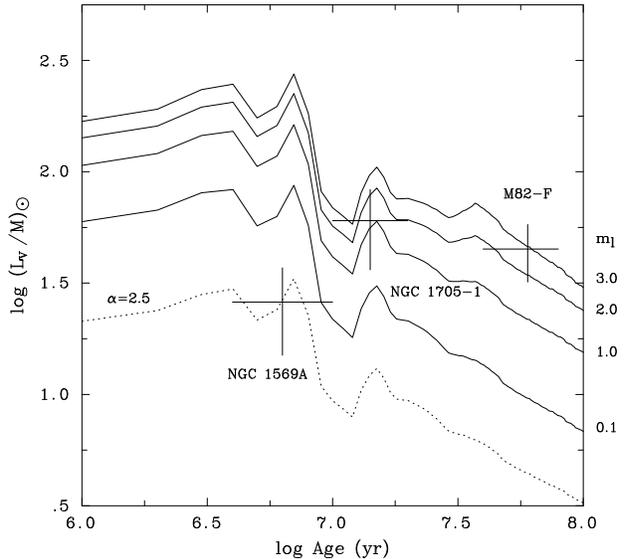}
\caption{($L_V/M)_\odot$ plotted as a function of age using the
Starburst99 code (Leitherer et al. 1999) taking into account mass-loss
due to stellar evolution. The solid curves shown are for solar
metallicity, an IMF parameterised by $\alpha=1.3$ for $0.1\le M
<0.5$~$M_\odot$ and $\alpha=2.3$ for $M \ge 0.5~$M$_\odot$, an upper
mass cut-off of 100~M$_\odot$, and lower mass cut-offs of 0.1, 1.0,
2.0 and 3.0~M$_\odot$.  The derived ($L_V/M)_\odot$ and age values for
M82-F, NGC 1569A and NGC 1705-1 are plotted. The dotted curve is for
solar metallicity, a single slope $\alpha=2.5$, and lower and
upper mass cut-offs of 0.1 and 100~M$_\odot$.}
\label{lm}
\end{figure}

To investigate the form of the mass function for M82-F and the two
other SSCs, we have calculated $(L_V/M)_\odot$ as a function of age
for various IMF parameterisations with the Starburst99 code.  Spectral
synthesis codes usually make the unrealistic assumption that the
mass-loss due to stellar evolution remains in the cluster.  In
calculating $(L_V/M)_\odot$, we have therefore allowed for a
decreasing cluster mass by subtracting the stellar ejecta at each time
step.

We adopt the parameterisation of Kroupa (2001) to represent our IMF
model: $\alpha=1.3$ for masses in the range $0.1\le M <0.5$~$M_\odot$
and $\alpha=2.3$ for $M \ge 0.5~$M$_\odot$ with an upper
mass limit $m_{\rm u}=100$~M$_\odot$, and lower mass limits in the
range $m_{\rm l}=$0.1--3~M$_\odot$. These models are shown in
Fig.~\ref{lm}.  It is clear that there is a fundamental mismatch by a
factor of $\sim 5$ between the derived ($L_V/M)_\odot$ for M82-F and
that predicted by a `simple stellar population' (SSP) model with a
standard Kroupa (2001) IMF with $m_{\rm l}=0.1$~M$_\odot$.

Both NGC 1569A and NGC 1705-1 have metallicities below solar
(Table~\ref{tab_comp}). We have investigated the effect of a lower
metallicity on ($L_V/M)_\odot$ by running the Starburst99 code for
$Z=0.2$ and 0.4$Z_\odot$. We find that the difference in
($L_V/M)_\odot$ is well within the observational errors because at
lower metallicities and young ages, the luminosity is increased but
the mass loss rate is lower.  Sternberg (1998) finds that NGC 1569A
has a steep MF with $\alpha \sim 2.5$, extending to
0.1~M$_\odot$. This is confirmed in Fig.~\ref{lm} where we show
($L_V/M)_\odot$ as a function of time for the Sternberg (1998) IMF
parameters.  For NGC 1705-1, Sternberg (1998) finds that the MF is
either flat or truncated. Our revised ($L_V/M)_\odot$ for this cluster
is lower by a factor of 2 but it still lies well above the IMF curve
extending down to $m_{\rm l}=0.1$~M$_\odot$ (Fig.~\ref{lm}),
indicating that NGC 1705-1 probably has a truncated MF with $m_l
\approx 1$--3~M$_\odot$.

The high visual luminosity-to-mass ratio that we have derived for
M82-F indicates that this SSC has an abnormal MF with a minor
fraction of its stellar mass concentrated in long-lived, low mass
stars. To explore this possibility, we
change the mass distribution by raising the lower cut-off mass in the
SSP models until we fit the observed ($L_V/M)_\odot$ ratio for
M82-F. As shown in Fig.~\ref{lm}, we need $m_{\rm l}=$2--3~M$_\odot$.
A similar lower mass range is indicated for a flatter IMF with a slope
of 2.0.  This exercise then implies a `top-heavy' MF where most of
the stellar mass is in stars with M$>$ 2~\Msun in the tightly
gravitationally bound regions of the cluster.

This conclusion merits careful examination; previous claims for highly
abnormal MFs often have proven incorrect.  Scalo (1998) summarized
the evidence for variations in the IMF. A key point from Scalo's
review is that most documented IMF variations seem to occur within
individual star-forming regions.  On large scales, the present-day IMF
seems to be remarkably constant (e.g. Kroupa 2001), and this trend
appears to extend to the MFs of the nearby SSCs NGC~3603 and R136a
(see discussion in Sect.~\ref{intro}). Thus it is not clear whether
M82-F represents a high- and intermediate-mass star-enriched core of a
larger star-forming complex, within which the overall MF is
relatively normal (see Sect.~\ref{formn}).

Some super star clusters appear to have formed as part of a more
extended region, e.g. knot S in the `antennae' interacting galaxies
(Whitmore et al. 1999) or the `nucleus' of the peculiar irregular
galaxy NGC~4449 (Gelatt, Hunter \& Gallagher 2001; B\"oker et
al. 2001), while in other cases, such as NGC~1569A, the SSC is well
separated from its surroundings (Hunter et al. 2000), and has a normal
MF (Fig.~\ref{lm}; Sternberg 1998).

The Galactic centre Arches and Quintuplet star clusters provide the
most compelling local examples of MF anomalies. Figer et al. (1999)
find both of these compact, moderate mass ($\sim 10^4$ \Msun) systems
have peculiarly flat MFs for M$>10$~M$_\odot$. They attribute this
behaviour to the strong tidal field near the Galactic centre that
could inhibit the formation of lower mass stars. 

Or is M82-F an example of an abnormal mode of star formation of the
type suggested from Kroupa's (2001) anaylsis of Galactic globular
clusters? With respect to this point, we can see from Fig.~\ref{lm} that an
SSP model with a single IMF is unlikely to fit all three SSCs with
mass determinations.  A larger sample of SSCs with mass and size
determinations is needed to resolve properly this issue. Until we do
so it is premature to generalize our result to the overall pattern of
star formation in M82.

\subsection{Formation and Mass Segregation}\label{formn}

Tightly bound star clusters are now widely recognized to be products
of locally efficient star formation; typically half or more of the gas
in the initial cloud's densest region must be converted into stars to
yield these types of objects (e.g. Goodwin 1997). M82-F must have
originated in a molecular cloud core with a density of
$n\sim10^5$~cm$^{-3}$ and a very high pressure (Elmegreen \& Efremov
1997). The discussion of these points by Ho and Filippenko (1996b) on
the connection between massive SSCs and ultra-dense molecular clouds
in M82 is borne out by our observations.

The high peculiar velocity of M82-F then is interesting. At present
neither dense molecular gas nor H\,{\sc II} emission are found at the
cluster's velocity, blueshifted by almost 170~km~s$^{-1}$ with respect
to the centroid of M82. The cluster evidently has had time to
completely separate itself from any natal gas. This is consistent with
our finding that the cluster is old enough to have made more than one
revolution about the centre of M82; it is now well-removed from its
birthplace and faces an uncertain future.

As discussed in Sect.~\ref{lum_r}, M82-F is too young to
have experienced significant dynamical mass segregation.
Some star clusters, however, seem to have been born with their
massive stars centrally concentrated.  The Orion nebula cluster sets
an example of this phenomenon; massive stars are centrally located but
the cluster is far too young for two-body relaxation to have played a
role (Hillenbrand \& Hartmann 1998).  Another example, NGC~2098 in the
LMC, has a similar age to M82-F, and shows some signs of mass
segregation (Kontizas et al. 1998). Fischer et al. (1998) reach a
similar conclusion for the 100~Myr old LMC star cluster NGC~2157.
Thus, if mass-segregation is playing a significant role in M82-F, it
must have been present at birth.

M82-F should therefore have an extended halo of low mass stars if
primordial mass segregation has occurred. To test for this
possibility, we take an extreme model where the velocity dispersion is
the same and assume that low mass stars affect the mass and not the
luminosity. The half-light radius has then to increase by a factor of
5 to 14~pc for $(L_V/M)_\odot$ to agree with a Kroupa-type IMF.  For
comparison, in Fig.~\ref{hst_F_BI}, clusters F and L are $\approx
70$~pc apart.  This increase is well beyond the radius where we can
detect M82-F, and while not impossible, there is no evidence at
present to support such an extended halo.  The other alternative is
that M82-F has had its halo of low mass stars removed by a process
such as tidal stripping. This mechanism has been proposed to explain
the peculiar mass functions of the Galactic centre clusters (Gerhard
2001). While M82-F does have a peculiar orbit, it seems
unlikely that it could have been born in a location with large tidal
stresses because such an environment is not obviously present in M82.
The data for the two younger SSCs NGC 1569A and 1705-1 are
inconclusive. The former cluster appears to have a normal MF while the
latter has a truncated MF yet both clusters reside in the inner parts
of their galaxies.

We conclude that while mass segregation at birth may have occurred in
M82-F, the current observational evidence favours the view that the
IMF of M82-F was abnormal in being deficient in low mass
stars. Future spatially-resolved spectroscopy of M82-F is needed to
investigate the significance of mass segregation.

\subsection{Long Term Survival?}\label{survival}
Predicting the lifetime of a star cluster is difficult. The simplest
model assumes a cluster is in virial equilibrium and therefore must
expand in response to any mass loss. In the case of M82-F this will be
dominated by effects due to stellar evolution.  At the age of M82-F,
the SSP models with lower mass cut-offs of 2 and 3~M$_\odot$,
show that 49 and 60 per cent of the original mass has been ejected.
Therefore, the cluster should soon have
too little mass to remain bound with a constant stellar velocity
dispersion; i.e. it will reach positive total energy.  How long the cluster
can endure depends on the actual form of the IMF, but since we expect that
less than half of the original stellar mass resides in stars with
M$<$2~M$_\odot$, the cluster should begin to dissolve in the next 1--2~Gyr as
stars in this mass range complete their evolution. We therefore adopt a
probable lifetime for the SSC M82-F in its present massive form of
$\leq2$~Gyr.

Of course star clusters are subject to a variety of internal and external
evolutionary influences beyond those imposed by stellar evolution and the
IMF, including tidal radius limits, stellar mass segregation, and the
nature of the stellar velocity distribution function (see Chernoff \&
Shapiro 1987).  The general impact of these additional influences is to
promote cluster expansion and thereby reduce the cluster lifetime, since
expansion causes increased vulnerability of stars to tidally-induced mass
loss (Spitzer 1987; Chernoff \& Weinberg 1990).  

A cluster's long term prospects further depend on its orbit, and the
importance of dynamical heating of its stars by processes such as disc
shocking (Gnedin \& Ostriker 1997).  Models of dense star clusters
therefore generally require a steep IMF extending to masses well below
that of the Sun to insure cluster survival over cosmic time scales.
For clusters with lower mass limits of $\sim$0.1~M$_\odot$, IMF slopes
of $\alpha \geq$2.3-2.5 are preferred for durable clusters in the
Milky Way, and clusters with $\alpha \leq1.5$ will disrupt even when
the tidal field is weak, as in the Magellanic Clouds (e.g.  Chernoff
\& Weinberg 1990).  Thus again, we are left with the question of
whether the ancient globular clusters, and especially those near the
dense centres of giant galaxies, were formed by some process
that was preferred at early epochs (see Zhang \& Fall 1999)?
Alternatively, it may simply be that the survivors were picked out from
what was initially a varied population of SSCs (e.g. Gnedin \& Ostriker
1997).

The large radial velocity of M82-F 
with respect to its host galaxy implies that M82-F cannot be on a simple
circular orbit within the disc of M82. It may be subject to some degree of
disc shocking as well as a moderate tidal radius constraint of about 35~pc
produced by M82. This is sufficiently large that tides should not have
an immediate impact on the structure of M82-F. However, if the cluster
begins to expand due to mass loss, as expected within the next 2~Gyr, then
tides will help to disperse the escaping stars.

\section{Conclusions}
We measure a stellar radial velocity dispersion of
13.4$\pm$0.7~km~s$^{-1}$ for the super star cluster M82-F. This result
in combination with a new determination of the projected half-light
radius of $r_{\rm h}=$2.8$\pm$0.3~pc yields a dynamical mass estimate
of $1.2\pm0.1 \times 10^6$~M$_\odot$ for M82-F. WFPC2 archival
observations together with a revised estimate for the minimum amount
of interstellar obscuration give $M_V = -14.5\pm0.3$ or $L_V =
5.4\pm1.4 \times 10^7$~L$_\odot$. The luminosity-to-mass ratio is then
$(L_V/M)_\odot = 45\pm13$. These parameters show that M82-F has the
luminosity and mass to qualify as a young globular star cluster.

Unlike most optically luminous SSCs, M82-F is middle-aged; we
reconfirm our earlier age estimate of 60$\pm$20~Myr from Paper I based
on fits to the H$\delta$ and He~I $\lambda$4922 absorption line
profiles.  {\it M82-F is a factor of 5 times too luminous for a star
cluster of its age with a standard stellar mass function extending to
0.1~M$_\odot$. Our data require that M82-F has a deficiency of low
mass stars, at least within the optically bright component of the
cluster.}

M82-F appears to be unique for an object of its age in displaying the
enhanced $(L_V/M)_\odot$ predicted for top-heavy IMFs that are
deficient in low mass stars. While this type of behaviour is seen in
Galactic centre clusters for M$>10$~M$_\odot$, it is not yet known if
this behaviour applies to lower stellar masses within these star
clusters, or if such objects should be expected to exist outside of
the very dense centres of giant galaxies.  Our models suggest that the
lower mass limit for M82-F is 2--3~M$_\odot$ for an IMF slope of 2.3,
and much of the mass is then near the lower stellar mass cutoff.  

We have considered obvious sources of error but have found none that
would substantially change our conclusions.  In particular, we have
investigated the question of mass segregation and find no current
observational evidence to support this possibility.  While modern
forms of the IMF which have fewer low mass stars than a Salpeter IMF
provide better agreement with our observed $(L_V/M)_\odot$ ratio, they
do not help with the dynamical problem of having sufficient low mass
stars to hold the cluster together over cosmic time spans. In this
case, the M82-F cluster is steadily losing mass due to mass lost by
evolving stars, and should dissolve within the next 1--2~Gyr. {\it If
the cluster is limited to what we see, its high mass and density are
not sufficient to assure its survival without sufficient mass in the
form of long-lived, low mass stars. M82-F cannot then be considered as
a proto-globular cluster.}

The history of evidence for substantial variations in the IMFs of star
clusters is chequered. For example, Elson, Fall \& Freeman (1989)
suggested that intermediate mass stars have a flat IMF in some LMC
star clusters, a view that gained support from an observed range in
cluster core radii consistent with expansion due to mass loss in the
presence of a flat IMF (Elson et al. 1989). However, in one of her
last papers, Elson et al. (1999) present deep STIS photometry of two
LMC star clusters that show rather normal MFs in both cases, even
though one cluster has a compact core and the other a large core of
the type previously associated with a flat IMF. In his review of IMFs,
Scalo (1998) suggests that significant IMF variations may occur
between star clusters, but finds no signs of IMF variations at the
level we require in M82-F.  Similarly, as we discussed earlier in
Sect.~\ref{intro}, neither of the nearest SSCs, NGC~3603 and R136,
have abnormal intermediate mass stellar MFs.

Further afield we re-examined properties of NGC~1569A from new
structural information published by Hunter et al. (2000) and in
NGC~1705-1 from size and photometric measurements taken from WFPC2
archival exposures for a new distance estimate (Tosi et al. 2001, in
prep.). We confirm the features discussed by Sternberg (1998) for
NGC~1569A which is well fit by SSP models with a slightly steeper
IMF with a single slope of 2.5. The new data for NGC~1705-1 also
confirm the results of Sternberg (1998) that this cluster has a
top-heavy MF; for a $\alpha=2.3$ slope, we find that the MF is truncated
near 2~M$_\odot$.

An indirect factor in favour of a possible top-heavy MF in M82-F is
the long held suspicion that this type of situation might generally
prevail in M82. Bernl\"ohr (1992) and Rieke et al. (1993) suggested a
deficit of low mass stars from fits to the spectral energy
distribution. Unfortunately, the luminosity of M82 depends critically
on assumptions concerning the degree of interstellar
obscuration. Satyapal et al. (1997) find that these earlier studies
over-corrected for interstellar obscuration, leading to an
artificially high stellar power output and the need for a top-heavy
MF. Doane and Mathews (1993) took a different approach. In modelling
the power requirements for the superwind, they needed a high supernova
rate that could be naturally explained by a top-heavy IMF.  Again more
modern approaches suggest that a normal IMF might be sufficient
(Strickland \& Stevens 2000).

From what we presently know, the over-luminous cluster M82-F is an
anomaly even within the anomalous M82 starburst. The only comparable
cases may be those seen in NGC~1275 SSCs by Brodie et al. (1998), who
presented spectroscopic evidence for top-heavy MFs. Yet theory
suggests that if IMF variations occur, they are most likley to appear
in dense star clusters, where mergers could take place and increase
the mass where the stellar IMF breaks in slope (Larson 1999). The SSC
M82-F then may be the long sought example of an evolved system with a
truly peculiar MF (see Kroupa 2001), and its properties and
environment therefore merit further careful investigation.
\section*{Acknowledgments}
We thank Monica Tosi for providing the distance to NGC 1705-1 in
advance of publication, and Richard Norris for his generous help with
the Starburst99 code. We also thank John Mathis for his comments on
this paper.  Both authors thank the organisers of the Massive Stellar
Clusters workshop held at the Observatory of Strasbourg in 1999 for a
very informative and stimulating meeting where many of the ideas
presented here were discussed.  JSG similarly expresses his
appreciation to participate in the Modes of Star Formation workshop
that took place at the Max Planck Institut f\"ur Astronomie in
Heidelberg during October 2000.  JSG thanks the Vilas Trustees for a
Vilas Associate through the University of Wisconsin-Madison Graduate
School. Key support was provided by the U.S. National Aeronautics and
Space Administration and Space Telescope Science Institute for studies
of massive star clusters through research funds associated with Hubble
Space Telescope observing programs, most recently as part of General
Observer Program 7446.  LJS thanks the Department of Astronomy at the
University of Wisconsin-Madison for their warm hospitality and
financial support during the writing of this paper.  The William
Herschel Telescope is operated on the island of La Palma by the Royal
Greenwich Observatory in the Spanish Observatorio del Roque de los
Muchachos of the Instituto de Astrof\'\i sica de Canarias.  The WIYN
Observatory is a joint facility of the University of
Wisconsin-Madison, Indiana University, Yale University and the
National Optical Astronomy Observatories.

\bsp
\label{lastpage}

\begin{thebibliography}{99}
\bibitem{}
Achtermann J. M., Lacy J. H. 1995, ApJ, 439, 163
\bibitem{}
Ashman K.M., Conti A., Zepf S.E., 1995, AJ, 119, 1164
\bibitem{}
Bica E., Santos J.F.C., Alloin D.,1990, Revista Mexicana de Astronomia y 
Astrofisica, 21, 202 
\bibitem{}
Bernl\"ohr K., 1992, A\&A, 263, 54 
\bibitem{}
Binney J., Tremaine S., 1987, Galactic Dynamics, 
Princeton University Press,  Princeton
\bibitem{}
B\"oker T., van der Marel R. P., Mazzuca L., Rix H.-W., Rudnick G.,
Ho L. C., Shields J. C., 2001, AJ, 121, 1473
\bibitem{}
Brodie J. P., Schroder L. L., Huchra J. P., Phillips A. C., 
Kissler-Patig M., Forbes D. A. 1998, AJ, 116, 691
\bibitem{}
Calzetti D., Kinney A. L., Storchi-Bergmann T. 1994, ApJ, 429, 582
\bibitem{}
Chernoff, D. F., Shapiro, S. L. 1987, ApJ, 322, 113
\bibitem{}
Chernoff D.F., Weinberg M.D., 1990, ApJ, 351, 121
\bibitem{}
De Marchi G., Clampin M., Greggio L., Leitherer C., Nota A., Tosi M., 1997,
ApJ, 479, L27
\bibitem{}
Devost D., Roy J.-R., Drissen L., 1997, ApJ, 482, 765
\bibitem{}
Doane J.S., Mathews W.G., 1993, ApJ, 419, 573
\bibitem{}
Dubath, P. \& Grillmair, C. J. 1997, A\&A, 321, 379
\bibitem{}
Eisenhauer F. et al 1998, ApJ, 498, 278
\bibitem{}
Elmegreen B., Efremov Y., 1997, ApJ, 480, 235
\bibitem{}
Elson R. A., Fall S. M., Freeman K. C. 1987, ApJ, 323, 54
\bibitem{}
Elson R. A., Fall S. M., Freeman K. C. 1989, ApJ, 336, 734
\bibitem{}
Elson R. A., Freeman K. C., Lauer T. R. 1989, ApJ, 347, L69
\bibitem{}
Elson R., Tanvir N., Gilmore G., Johnson R. A., Beaulieu, S. 1999, 
in Chu Y.-H., Suntzeff N. B., Hesser J. E., Bohlender D. A., eds.,
IAU Symp. No. 190. New Views of the Magellanic Clouds,  
Astron. Soc. Pac., San Francisco p. 417
\bibitem{}
Figer D. F., Kim S. S., Morris M., Serabyn E., Rich R. M., McLean I.
S., 1999, ApJ, 525, 750
\bibitem{}
Fioc M., Rocca-Volmerange B., 1997, A\&A, 326, 950
\bibitem{}
Fischer P., Pryor C., Murray S., Mateo M., Richtler T., 1998, 
AJ, 115, 592
\bibitem{}
Freedman W. et al., 1994, ApJ, 427, 628
\bibitem{}
Fritze - v.  Alvensleben U., 1998, A\&A, 336, 83
\bibitem{}
Fritze - v. Alvensleben U., 1999, A\&A, 342, L25
\bibitem{}
Gallagher J.S., III, Smith L.J., 1999, MNRAS, 304, 540
\bibitem{}
Gelatt A. E., Hunter D. A., Gallagher J. S., 2001, PASP, 113, 142
\bibitem{}
Gerhard O., 2000, in Lan\c{c}on A., Boily C.M., eds., ASP Conf. Ser. Vol. 211,
Massive Stellar Clusters. Astron. Soc. Pac., San Francisco p. 12
\bibitem{}
Gerhard O., 2001, ApJL, 546, L39
\bibitem{}
Gnedin O. Y., Ostriker J. P. 1997, ApJ, 474, 223
\bibitem{}
Gonz\'alez Delgado R.M., Leitherer C., 1999, ApJS, 125, 479
\bibitem{}
Gonz\'alez Delgado R.M., Leitherer C., Heckman T.M., 1999, ApJS, 125, 489
\bibitem{}
Gonz\'alez Delgado R.M., Leitherer C., Heckman T.M., Cervi\~{n}o M., 1997,
ApJ, 483, 705
\bibitem{}
Goodwin S. P. 1997, MNRAS, 286, 669
\bibitem{}
G\"otz M., McKeith C. D., Downes D., Greve A. 1990, A\&A, 240, 52\bibitem{}
Gray D.F., Toner C.G., 1986, ApJ, 310, 277
\bibitem{}
de Grijs R., O'Connell R. W., Becker G. D., Chevalier R. A.,
Gallagher J. S., III, 2000, AJ, 119, 681
\bibitem{}
de Grijs R., O'Connell R. W., Gallagher J. S., III, 2001, AJ, 121, 768
\bibitem{}
Harris W. E. 1991, ARA\&A, 29, 543
\bibitem{}
Heckman T.M., Leitherer C., 1997, AJ, 114, 69
\bibitem{}
Hillenbrand L.\ A., Hartmann L.\ W., 1998, ApJ,  492, 540 
\bibitem{}
Ho L.C., Filippenko A.V., 1996a, ApJ, 466, L83
\bibitem{}
Ho L.C., Filippenko A.V., 1996b, ApJ, 472, 600
\bibitem{}
Holtzman J.A., Burrows C.J., Casertano S., Hester J.J., Trauger J.T., 
Watson A.M., Worthey G., 1995, PASP, 107, 1065
\bibitem{}
Hunter D.A., O'Connell R.W., Gallagher J.S., Smecker-Hane Y.T., 2000,
AJ, 120, 2383
\bibitem{}
Jacoby G.H., Hunter D.A., Christian C.A., 1984, ApJS, 56, 257
\bibitem{}
Kim S.S., Morris M., Lee H.M., 1999, ApJ, 525, 228
\bibitem{}
Kobulnicky H.\ A., Skillman E.\ D., 1997, ApJ,  489, 636 
\bibitem{}
Kontizas M., Hatzidimitriou D., Bellas-Velidis I., Gouliermis D., 
Kontizas E., Cannon R. D., 1998, A\&A, 336, 503
\bibitem{}
Kroupa P. 2001, MNRAS, 322, 231
\bibitem{}
Larson R.B., 1985, MNRAS, 214, 379
\bibitem{}
Larson R. B. 1999, in T. Nakamoto, ed., Star Formation 1999, Nobeyama
Radio Observatory, p.336-340,
\bibitem{}
Leitherer C. et al., 1999, ApJS, 123, 3 
\bibitem{}
Lester D. F., Carr J. S., Joy M., Gaffney N. 1990, ApJ, 352, 544
\bibitem{}
Massey P., Hunter D. A. 1998, ApJ, 493, 180
\bibitem{}
McCleod K.K., Rieke G.H., Rieke M.J., Kelly D.M., 1993, ApJ, 412, 111
\bibitem{}
Meurer G.R., Heckman T.M., Leitherer C., Kinney A., Robert C.,
Garnett D.R., 1995, AJ, 110, 2665
\bibitem{}
O'Connell R.W., Mangano J.J., 1978, ApJ, 221, 62
\bibitem{}
O'Connell R.W., Gallagher J.S., III, Hunter D.A., 1994, ApJ, 433, 65
\bibitem{}
O'Connell R.W., Gallagher J.S., III, Hunter D.A., Colley W.N. 
1995, ApJ, 446, L1
\bibitem{}
Phillips A.C., Guzman R., Gallego J., Koo D.C., Lowenthal J.D., Vogt N.P., 
Faber S.M., Illingworth, G.D., 1997, ApJ, 489, 543
\bibitem{}
Rieke G.H., Lebofsky M.J., Thompson R.I., Low F.J., Tokunaga A.T.,
1980, ApJ, 238, 24
\bibitem{}
Rieke G.H., Loken K., Rieke M.J., Tamblyn P., 1993, ApJ, 412, 99
\bibitem{}
Sait\={o} M., Sasaki M., Kaneko N., Nishimura M., Toyama K., 1984, 
PASJ, 36, 305
\bibitem{}
Sakai S., Madore B.F., 1999, ApJ, 526, 599
\bibitem{}
Satyapal S. et al., 1995, ApJ, 448, 611
\bibitem{}
Satyapal S., Watson D.M., Pipher J.L., Forrest W.J., Greenhouse M.A., 
Smith H.A., Fischer J., Woodward C.E., 1997, ApJ, 483, 148
\bibitem{}
Scalo J. 1998, in Gilmore G., Howell, D., eds., in ASP
Conf. Ser. Vol. 142. The Stellar Initial Mass Function,
Astron. Soc. Pac., San Francisco p. 201
\bibitem{}
Schiavon R.P., Barbuy B., Bruzual A.G., 2000, ApJ,  532, 453 
\bibitem{}
Schweizer F., Seitzer P. 1993, ApJ, 417, L29
\bibitem{}
Seaquist E.R., Carlstrom J.E., Bryant P.M., Bell M.B., 1996, ApJ, 465, 691
\bibitem{}
Shen J., Lo K. Y. 1995, ApJ, 445, L99
\bibitem{}
Shortridge K., Meyerdierks H., Currie M., Clayton M.,
1997, Starlink User Note 86.13, Rutherford Appleton Laboratory
\bibitem{}
Sirianni M., Nota A., Leitherer C., De Marchi G., Clampin, M. 2000, 
ApJ, 533, 203
\bibitem{}
Sofue Y., Reuter H.-P., Krause M., Wielebinski R., Nakai N. 1992,
ApJ, 395, 126 
\bibitem{}
Spitzer L., Jr. 1987, Dynamical Evolution of Globular Clusters,
Princeton Univ. Press, Princeton
\bibitem{}
Spitzer L., Jr., Hart M.H. 1971, ApJ, 164, 399
\bibitem{}
Sternberg A., 1998, ApJ, 506, 721
\bibitem{}
Strickland D. K., Stevens I. R. 2000, MNRAS, 314, 511
\bibitem{}
Takahashi K., Portegies Zwart S.F., 2000, ApJ, 535, 759
\bibitem{}
Telesco C. M. 1988, ARA\&A, 26, 343
\bibitem{}
Whitmore B.C., Schweizer F., Leitherer C., Borne K., Robert C.,
1993, AJ, 106, 1354
\bibitem{}
Whitmore B.C., Zhang Q., Leitherer C., Fall S.M., Schweizer F., Miller B.W.,
1999, AJ, 118, 1551
\bibitem{}
Yun M. S., Ho P. T. P., Lo K. Y. 1993, ApJ, 411, L17
\bibitem{}
Zhang Q., Fall S. M. 1999, ApJ, 527, L81
\end{thebibliography}
\end{document}